# Anti-chiral edge states based on photonic Floquet lattices

**Junying Wang**[1,#], **Xifeng Ji**[2, #], **Zhiwei Shi**[2,*], **Yajing Zhang**[2], **Huagang Li**[3], **Yang Li**[2], **Yaohua Deng**[2,†], **Kang Xie** [2,‡]

1  School of Information Engineering, Guangdong University of Technology, Guangzhou 510006, China
2  School of Electromechanical Engineering, Guangdong University of Technology, Guangzhou 510006, China
3  School of Photoelectric Engineering, Guangdong Polytechnic Normal University, Guangzhou 510665, China
# Junying Wang and Xifeng Ji contributed equally to this paper
Correspondence: ∗szwstar@gdut.edu.cn; †dengyaohua@gdut.edu.cn; ‡kangxie@gdut.edu.cn

**Abstract:** Photonic Floquet lattices provide an excellent platform for manipulating different topologically protected edge states. However, anti-chiral edge states have not been discussed much in Floquet lattices. Here, we propose a waveguide structure by combining two honeycomb Floquet photonic lattices with opposite rotation directions. In this structure, we find that the anti-chiral edge states have the same transmission direction on two parallel body edges. With an increasing modulation phase difference between the two sublattices in one direction, the width of the band gap becomes smaller and the robustness of the edge states becomes weaker. Interestingly, the transmission speed is also controlled by the phase difference. In addition to their relevance for the topological properties of the Floquet lattice system, these results may be applied to multi-channel optical switches, optical functional devices, and in other fields.
**Keywords**: photonic Floquet lattices; anti-chiral edge states; phase difference

**Introduction**

In the past few decades, inspired by the discovery of the quantum Hall effect and topological insulators in condensed-matter physics, scientists have compared the electrons in condensed-matter physics systems with photons in photonic systems, extended the concept of controlling material properties to the optical field, and created topological photonics[1,2], which have enabled the rapid rise and development of this new frontier field via, for example, the one-dimensional (1D) Su-Schrieffer-Heeger (SSH) model[3–5], 2D optical integer quantum Hall effect[6–8], optical quantum spin Hall effect[9–11], optical Floquet topological insulator[12–14], and 3D Weyl point[15–17] and nodal line[18–20]. The introduction of nonlinear effects[21–23], non-Hermite systems[24–26], valley states[27–29], and the discovery of high-order topological insulators[30–32] have brought more abundant physical properties to photonic systems, and also brought challenges in theory and design. On this basis, many interesting topological devices have also been developed and studied, such as topological lasers[33–35], optical delay lines[36–38], unidirectional waveguides[39–41], and 3D topological insulators[42].

Generally, two types of topologically protected edge states exist. One is the chiral topological transport state[6–8] in which the topologically protected edge transport mode must propagate clockwise or counterclockwise along the edge of the body. The other topologically protected edge state is a spiral topological transport state with a pair of edge transport states with opposite group velocities[9–11]. However, Colomes *et al*.[43] improved the Haldane model in 2018 by changing the jump parameters of the next-nearest neighbor in the sub-lattice, and the energy of two adjacent Dirac points shifted in the opposite direction; they further proposed the concept of anti-chiral edge states in 2D systems.

This interesting phenomenon has also been studied theoretically in many other physical systems, such as graphene structures combining a magnetic field and interlayer bias in graphene multilayer systems[44,45], exciton polarization bands of polarization-dependent interactions of polaron condensates[46], and honeycomb lattices composed of Heisenberg ferromagnets that break sub-lattice symmetry[47]. Because it is difficult to break the symmetry of time inversion, it is difficult to observe the anti-chiral edge states in reality. However, Zhou *et al*. [48] and Yang *et al*. [49] have recently proved and observed the existence of anti-chiral edge states in gyromagnetic photonic crystals and classical circuit lattices through experiments. Recently, Cheng *et al*. realized anti-chiral edge states theoretically by constructing a photonic crystal composed of two original Haldane-model subsystems with opposite chirality[50]. Thus, it may be imagined that the simplest way to realize anti-chiral edge states is to have more than two subsystems with different chiralities in a system.

By periodically driving to destroy the time-reversal symmetry of a system, new topological properties of materials can be obtained[51–57]. Rechtsman *et al*. broke the time-reversal symmetry by introducing a spiral waveguide array[53] and obtained a photonic Floquet topological insulator. Because of their different rotations, such insulators will generate chiral topological edge states in different directions. Thus, by combining two spiral waveguide Floquet systems with opposite rotation directions—that is, two Floquet subsystems with opposite chiralities[54,55]—the anti-chiral edge states may be naturally realized. In this paper, we prove this conjecture and find that anti-chiral edge states propagate in the same direction on two parallel edges in a strip geometry.

First, we propose a 3D Floquet lattice model[55] that is periodic in the $y$ direction and finite in the $x$ direction, as shown in Fig. 1(a). The model is composed of two regions (I and II) along the x-axis direction, all waveguides are characterized by equal helix radii $R$, the same longitudinal helix period $T$ is along the z axis, and all are separated by the distance $d$ in the transverse plane. Therefore, the model can be represented by a periodic optical potential $V = (x, y, z) = (x, y, z + T)$. This optical periodic potential $V$ can be expressed as[55]

$$V = \begin{cases} \sum_{m,n} V_0 \, exp\left(-\frac{[x-R\sin(\Omega z)-x_{m,n}]^2+[y+R\Omega\cos(\Omega z)-y_{m,n}]^2}{w^2}\right) & (x_{m,n} \leq 0) \\ \sum_{m,n} V_0 \, exp\left(-\frac{[x+R\sin(\Omega z+\Delta\varphi)-x_{m,n}]^2+[y-R\Omega\cos(\Omega z)-y_{m,n}]^2}{w^2}\right) & (x_{m,n} > 0) \end{cases},$$

(1)

where $V_0$ is the amplitude of potential, w is the waveguide diameter, $\Omega = 2\pi/T$ is the helix frequency, $\Delta\varphi$ is the modulation phase difference in the $x$ direction, and $m$ and $n$ are the specific positions of each helix waveguide. By setting the waveguide positions, we can build a cellular photonic lattice composed of two helix waveguides with opposite directions. Here, we assume that, in the Floquet photonic lattice, $a = 1$ is the normalized lattice constant, the distance between adjacent waveguides $d = a/\sqrt{3}$, and the waveguide radius $w = 0.2d$.

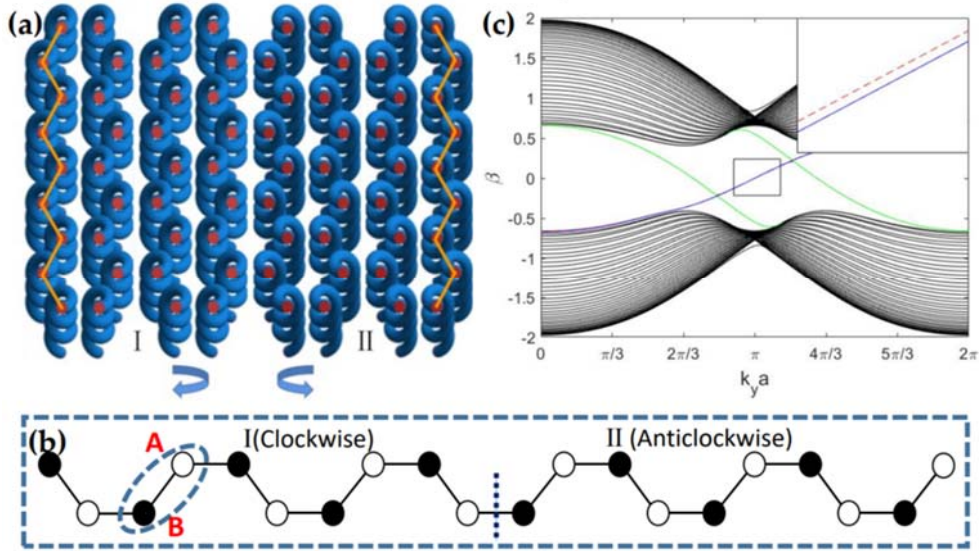

**Fig. 1. (a)** 3D lattice structure with opposite helicities. **(b)** Supercell of photonic Floquet lattices in (a) at $\Delta\varphi = 0$. **(c)** Dispersion curves of edge states with $R = 0.3d$ in (a). Inset is a magnified view of curve in the interval $[8\pi/9, 10\pi/9]$ to illustrate the existence of two anti-chiral edge states.

We constructed a supercell of a photonic Floquet lattice in Fig. 1(b), with the clockwise rotating helical waveguide on the left (I) and the anticlockwise rotating helical waveguide on the right (II). According to the tight-binding approximation, the Hamiltonian $H$ of this model can be described as[54]

$$H = \vec{t} \cdot \left[\sum_{m=1}^{N}(|m,B\rangle\langle m,A| + h.c.), \sum_{m=1}^{N-1}(|m+1,B\rangle\langle m,A| + h.c.), \sum_{m=1}^{N-1}(|m-1,B\rangle\langle m,A| + h.c.)\right], \quad (2)$$

where t is the coupling coefficient, $A$ and $B$ different waveguides in a single cell, and $N$ the number of $A$ or $B$. Thus, the dispersion curves of the system can be obtained, as shown in Fig. 1(c). The solid blue and dashed red lines represent the left- and right-hand edge states, respectively, in the supercell in Fig. 1(b), and their transmission directions are the same, which are anti-chirality edge states. The solid green line represents the edge states of the middle domain wall in Fig. 1(b), the transmission directions of which are opposite those of the edge states on both sides, which can be seen in Fig. 1(c).

Sub-lattices I and II of the system are composed of clockwise spiral waveguides and anti-clockwise waveguides, respectively. The topologically invariant named gap Chern number [55] for the clockwise helical waveguide is $C_{v1} = 1$, and the gap Chern number for the anticlockwise helical waveguide is s $C_{v2} = -1$. The local gap Chern numbers of the two sub-lattices have opposite signs, so the corresponding two edge states should have opposite chirality. Physically, the two boundaries (left- and right-hand boundaries) of the system have two edge states in the same direction of transmission, while the counterpropagating states should be located in the centrally transverse direction.

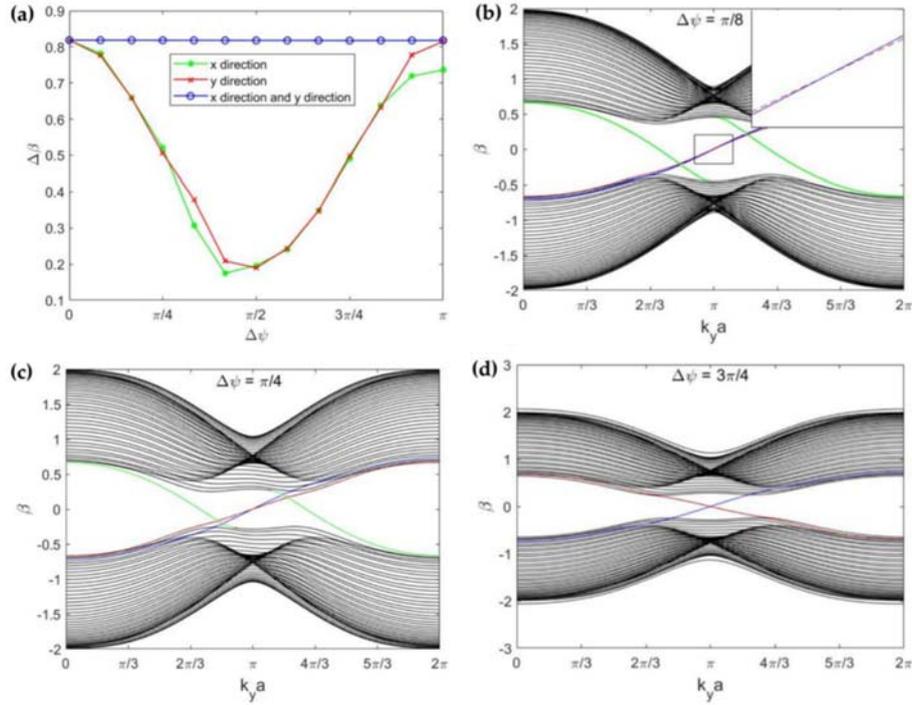

Fig. 2. **(a)** Relations of the band-gap width of the supercell in Fig. 1(b) with modulation phase difference when $R = 0.3d$. The green line indicates the relationship between band-gap width and modulation phase difference in the $x$ direction, the red line indicates the relationship between band-gap width and modulation phase difference in the y direction, and the blue line indicates the relationship between band-gap width and modulation phase difference synchronized in $x$ and $y$ directions. **(b), (c), and (d)** Dispersion curves of edge state when modulation phase difference is $\Delta\varphi = \pi/8$, $\pi/4$, and $3\pi/4$, respectively, where $R = 0.3d$. Inset in (b) shows a magnified view of the curve in the interval $[8\pi/9, 10\pi/9]$ to illustrate the existence of two anti-chiral edge states.

Next, we changed the modulation phase difference $\Delta\varphi$ in the $x$ direction. Owing to the particularity of the properties of trigonometric functions, $\Delta\varphi$ can only change in the interval from 0 to $2\pi$, and the change from 0 to $\pi$ is opposite the change from $\pi$ to $2\pi$. By calculating its dispersion curve, we find that when $\Delta\varphi$ increases in the range from 0 to $\pi/2$, the band-gap width of the system becomes smaller. At this time, the robustness of the anti-chiral edge states at x ≥ 0 becomes weaker, and the transmission speed becomes slower, as shown in Figs. 2(b)–2(c). With increasing Δ$\varphi$ in the range of $\pi/2$ to $\pi$, the band-gap width of the system becomes larger, the edge states at the middle domain wall disappear, and the anti-chiral edge states become chiral, as shown in Figs. 2(a) and 2(d). The reason for this is that the Floquet lattice is clockwise at $x \geq 0$, which is the same as the lattice at $x < 0$. In addition, we can change the modulation phase difference in the $y$ direction. The results are consistent with the above results in the $x$ direction. However, when we synchronously change the modulation phase difference in the $x$ and $y$ directions, it is found that the band-gap width of the system and the strength of the edge state is unchanged, and the synchronous phase-difference change has no effect on the anti-chiral edge states; see Fig. 2(a).

We further verified the existence and properties of the anti-chiral edge states by simulating beam propagation on the lattice structures. We launched two elongated Gaussian beams $\phi_1$ and $\phi_2$ into the left- and right-hand boundaries of the helical lattice. The expression of these Gaussan beams are

$$\phi_1 = \phi(x, y, z = 0) = A_0 \exp\left(\frac{-(x - x_1)^2}{\omega_x^2} - \frac{y - y_1}{\omega_y^2}\right) \exp(ik_y y),$$

$$\phi_2 = \phi(x, y, z = 0) = A_0 \exp\left(\frac{-(x - x_2)^2}{\omega_x^2} - \frac{y - y_2}{\omega_y^2}\right) \exp(ik_y y),$$

(3)

where $A_0$ is the amplitude, $\omega_x = 0.5a, \omega_y = 4a, k_y = \pi/a, x_1 = -19a, y_1 = 21a, x_2 = 19a$, and $y_2 = 21a$. It can be seen that the propagation directions of the left- and right-hand boundary beams of the photonic Floquet lattices are the same, and the transmission is stable without dispersion, as shown in Figs. 3(a)–3(f). That is to say, they are anti-chiral edge states. We then added a defect at the left-hand boundary $x = 13a, y = 16a$, on the basis of the original lattice [see Fig. 3(g)]; the lattice parameters are unchanged and the same two beams are used for transmission. Before $Z = 20T$, the propagation modes on both sides are the same. After $Z = 20T$, the light beam starts to travel to the defect [Fig. 3(h)] and continues to travel through the defect [Fig. 3(i)]. These all indicate that the chirality-resistant edge state of the system is very robust.

In addition, we increased the phase difference at system $x \geq 0$ for analog transmission, and the results verify that when the phase difference of $\pi/8$ is introduced in the x direction, the transmission effect is shown in Figs. 4(a) and 4(b). When the phase difference of $\pi/4$ is introduced in the x direction, the transmission effect is shown in Figs. 4(c) and 4(d). Here, we define the x-direction beam center ("center of mass") as $x_c = \iint x |\phi_2|^2 dxdy / \iint |\phi_2|^2 dxdy$. We find that, as the phase difference increases in the range $0 - \pi/2$, the robustness of the anti-chirality edge state at $x \geq 0$ becomes weaker, and the transmission speed becomes slower, as shown in Figs. 4(e) and 4(f), which is consistent with the results of the dispersion structure [Figs. 2(b) and 2(c)].

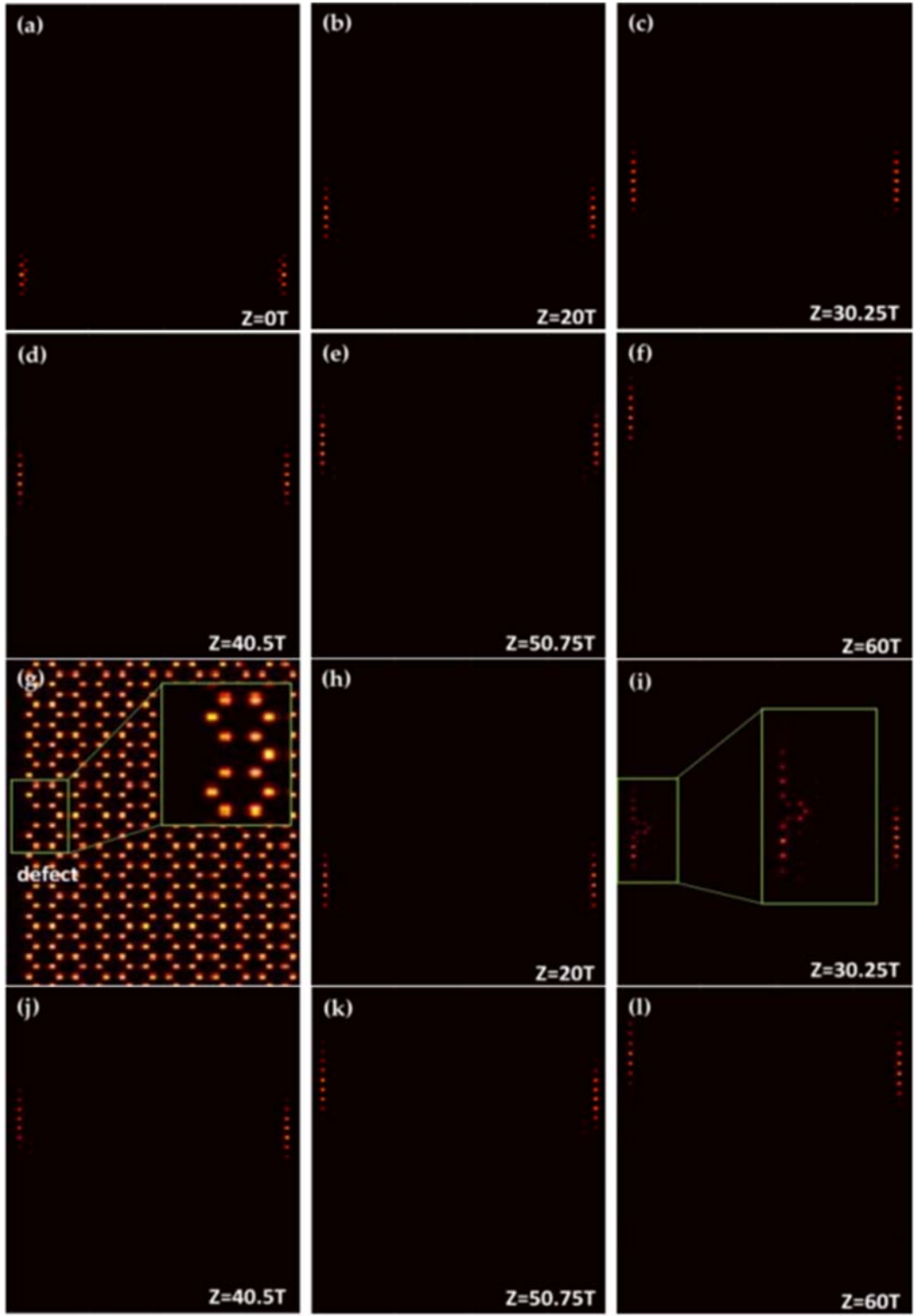

**Fig. 3.** (**a**)–(**f**) Propagation of topological edge state at different propagation distances. In (**g**), a defect is added to the lattice structure. (**h**)–(**l**) Propagation of topology edge state at different propagation distances after adding a defect.

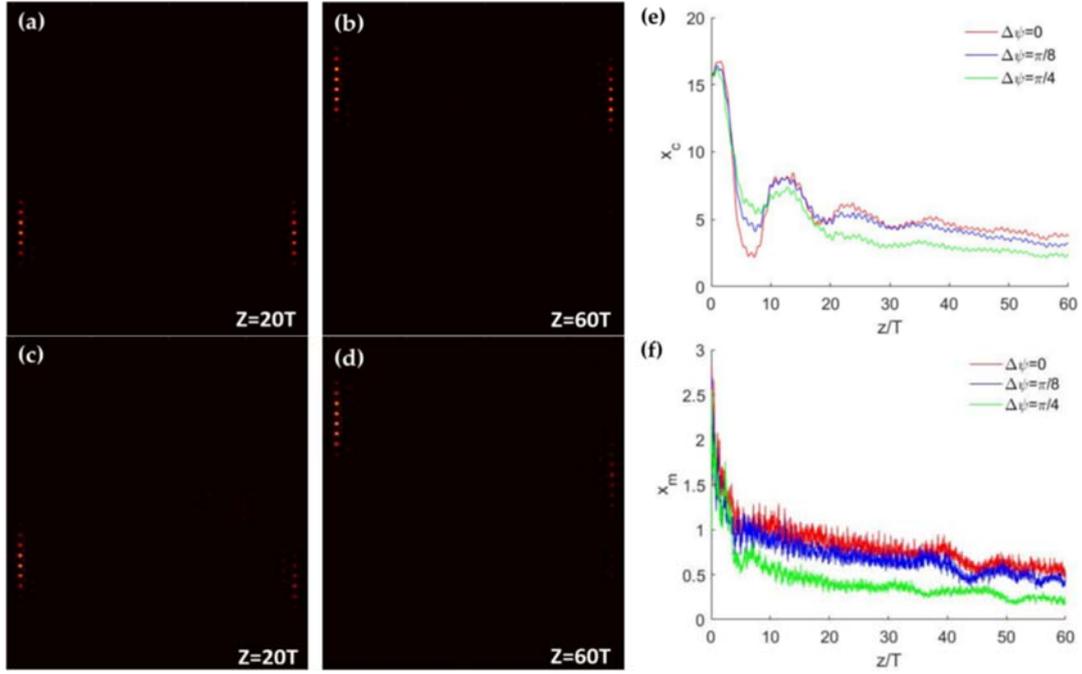

**Fig. 4.** (**a**) and (**b**) Transmission diagram when phase difference in the $x$ direction is $\pi/8$. (**c**) **and** (**d**) Transmission diagram when phase difference in the $x$ direction is $\pi/4$. (**e**) Beam center (centroid) in the x direction of $\phi_2$ as function of relative propagation distance $z/T$. (**f**) Position of peak beam power $x_m$ in the direction of $\phi_2$ as function of relative propagation distance $z/T$.

In summary, in this paper, we propose and prove that the system composed of two photon Floquet sub-lattices with opposite rotation directions can support topologically protected anti-chiral edge states. We find that by changing the modulation phase difference of two sub-lattices with opposite rotation directions, the band-gap widths and transmission speeds will both change. The stability, absence of dispersion, and strong robustness of the system anti-chiral edge states are further verified by simulated transmission experiments and the addition of defects. The theory of topologically protected anti-chiral structures has only been put forward in recent years in photonic systems, so these findings may lead to a new theoretical model in this field and may be applied to related multi-channel photonic devices.

Author Contributions are as follows: Conceptualization, Zhiwei Shi; methodology, Xifeng Ji, Yajing Zhang and Junying Wang; validation, Xifeng Ji, Junying Wang and Huagang Li; formal analysis, Yang Li and Yaohua Deng; investigation, Zhiwei Shi, Yaohua Deng, and Kang Xie; writing—original draft preparation, Xifeng Ji and Junying Wang; writing—review and editing, Zhiwei Shi and Kang Xie; funding acquisition, Yaohua Deng and Kang Xie. All authors have read and agreed to the published version of the manuscript.

Conflicts of Interest: The authors declare that they have no known competing financial interests or personal relationships that could have appeared to influence the work reported in this paper.

Acknowledgments: This work was supported by the Guangdong Basic and Applied Basic Research Foundation (2020A1515010623), and the Natural Science Foundation of China (11874126, 52175457).